\documentclass[prl,aps,preprint,floatfix,superscriptaddress]{revtex4-2}
\usepackage{amsmath,amssymb,amsfonts}
\usepackage{graphicx}
\usepackage{bm}
\usepackage{hyperref}

\begin{document}

\title{Generalized Critical-State Analysis of Local Field Screening in High-Pressure Yttrium Polyhydride}

\author{A.~G.~Gavriliuk}
\affiliation{Center for High Pressure Science and Technology Advanced Research, Shanghai, China}
\affiliation{Institute for Nuclear Research, Russian Academy of Sciences, Moscow, Troitsk, 108840, Russia}
\affiliation{Shubnikov Institute of Crystallography of the Kurchatov Complex Crystallography and Photonics, NRC ``Kurchatov Institute'', Moscow, 119333, Russia}
\affiliation{Immanuel Kant Baltic Federal University, Kaliningrad 236041, Russia}
\author{I.~A.~Troyan}
\affiliation{Institute for Nuclear Research, Russian Academy of Sciences, Moscow, Troitsk, 108840, Russia}
\affiliation{Shanghai Advanced Research in Physical Sciences, Shanghai, China}
\author{A.~A.~Mironovich}
\affiliation{Institute for Nuclear Research, Russian Academy of Sciences, Moscow, Troitsk, 108840, Russia}
\author{A.~G.~Ivanova}
\affiliation{Institute for Nuclear Research, Russian Academy of Sciences, Moscow, Troitsk, 108840, Russia}
\affiliation{A. V. Shubnikov Institute of Crystallography of the Kurchatov Complex Crystallography and Photonics, NRC ``Kurchatov Institute'', Moscow, 119333, Russia}
\author{H.-K.~Mao}
\affiliation{Shanghai Advanced Research in Physical Sciences, Shanghai, China}
\affiliation{Center for High Pressure Science and Technology Advanced Research, Shanghai, China}
\author{V.~V.~Struzhkin}
\affiliation{Center for High Pressure Science and Technology Advanced Research, Shanghai, China}

\date{\today}

\begin{abstract}
We resolve the long-standing controversy over the interpretation of magnetic-flux expulsion in high-pressure superhydrides by measuring the local magnetic field adjacent to yttrium polyhydride at 171~GPa and 4.1~K with a $^{57}$Fe M\"ossbauer sensor. Nuclear forward scattering tracks the local induction $B_{\rm sensor}(B_a)$, which nearly merges with the normal-state line near 1.5~T. Rather than assuming a constant-current Bean state, we solve the critical-state problem for the actual thin-disk geometry using the Shantsev--Clem--Sanchez method and compare generalized field-dependent critical-state models. Both the Kim and exponential laws reproduce the measured virgin branch, whereas the constant-$J_c$ Bean model fails. The corresponding zero-field critical-current densities, $(4.38\pm1.48)\times10^{11}$ and $(2.45\pm0.22)\times10^{11}$~A/m$^2$, respectively, are consistent in order of magnitude with transport estimates for granular YH$_6$. These results establish that the observed flux expulsion is a pinned critical state, not the equilibrium Meissner effect, and demonstrate a general methodology for probing local superconductivity under extreme conditions.
\end{abstract}

\maketitle

\section{Introduction}

The discovery of near-ambient superconductivity in polyhydrides under high pressure has stimulated a sustained effort to detect magnetic signatures of superconductivity---the Meissner effect, magnetic flux trapping, and hysteretic magnetization---using a variety of local and bulk magnetometry techniques. Following the initial report of superconductivity in H$_3$S at 203~K~\cite{Drozdov2015}, record transition temperatures were subsequently reported in LaH$_{10}$~\cite{Somayazulu2019,Drozdov2019} and related hydride systems. Minkov \textit{et al.} reported magnetic-field screening and magnetization measurements on sulfur hydride and lanthanum hydride under pressure~\cite{Minkov2022} and subsequently reported trapped-flux studies on the same systems~\cite{Minkov2023}. NV-center magnetometry has provided spatially resolved images of Meissner screening and flux trapping in CeH$_9$~\cite{Bhattacharyya2024} and LaH$_{10}$~\cite{Chen2026}. Nuclear forward scattering with an embedded M\"ossbauer sensor was first applied to a superhydride by Troyan \textit{et al.}~\cite{Troyan2016}, who reported field expulsion in H$_3$S. NFS is a well-established synchrotron-based technique for probing magnetism and hyperfine interactions in condensed matter \cite{Gerdau1985,Hastings1991,Wortmann2002,Lubbers1999,Sturhahn2004,Roehlsberger2004}, with multiple applications under extreme conditions~\cite{Gavriliuk2005,Gavriliuk2018,Gavriliuk2023,Li2006,Bi2015}.

The interpretation of the NFS experiment, in particular, has been contested. The original work on H$_3$S~\cite{Troyan2021} treated the data as evidence for the Meissner effect, whereas Hirsch and Marsiglio~\cite{Hirsch2021} argued that the observed screening is instead consistent with a Bean-type critical state~\cite{Bean1962} with a field-independent \(J_c\). Their critique is internally consistent as a conditional Bean-model argument---strong screening without observable trapped flux or magnetization consistency would be problematic---but it is not a direct falsification because geometry, granularity, and pinning assumptions strongly affect both predictions. Prozorov and Bud'ko~\cite{Prozorov2022} subsequently showed from finite-geometry calculations that neither a simple London--Meissner description nor the proposed nonstandard scenario naturally accounts for both field orientations, whereas a conventional type-II critical state with moderate pinning can do so. Their rebuttal supports the vortex-pinning interpretation, although the constant-current Bean approximation itself remains incomplete. Here we show that the constant-current description is not adequate. By solving the critical-state problem for the actual thin-disk geometry with the Shantsev--Clem--Sanchez (SCS) method~\cite{Clem1994,Shantsev1999}, which extends earlier analytical solutions for thin disks~\cite{Mikheenko1993,Zhu1993} to arbitrary field-dependent \(J_c(H)\), we find that two generalized critical-state models, the Kim law~\cite{Kim1962} \(J_c(H)=J_{c0}/(1+H/B_0)\) and the exponential law~\cite{Fietz1964,Xu1990} \(J_c(H)=J_{c0}\exp(-H/B_0)\), reproduce the measured ZFC (virgin) branch. The Bean model, by contrast, retains a high-field offset and fails. We therefore retain both field-dependent models as candidates and use their spread to estimate the physically relevant range of \(J_{c0}\). This resolves the Troyan--Hirsch controversy without invoking equilibrium Meissner physics, while also showing that a single virgin curve does not uniquely discriminate between different pinning laws. 

\section{Methods}

\begin{figure}[t]
\includegraphics[width=\columnwidth]{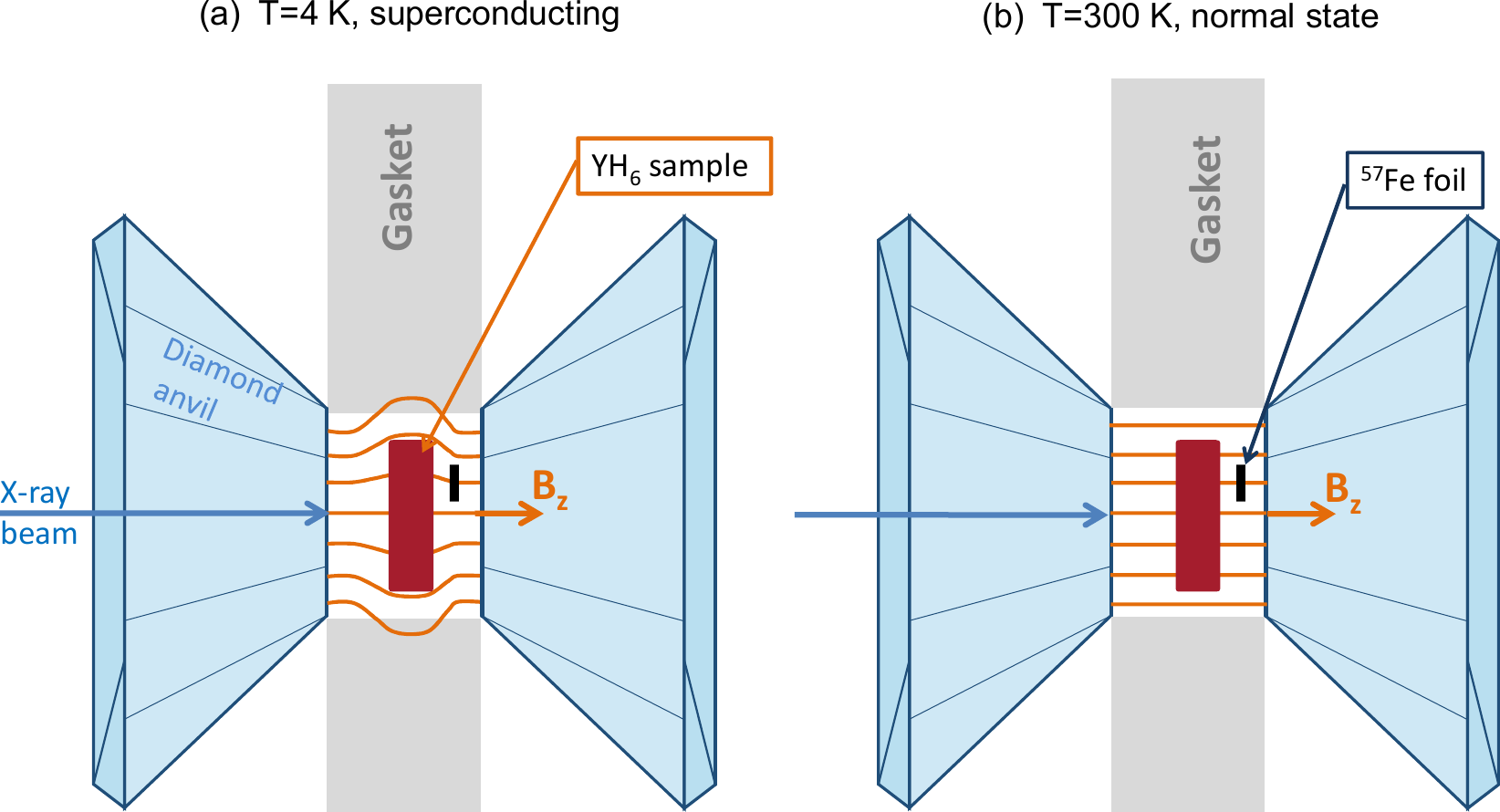}
\caption{Schematic of the NFS experiment on $^{57}$Fe nuclei at (a) $T=4$~K and (b) $T=300$~K. The sample-chamber dimensions are exaggerated to show the geometry; the initial thickness of the medium layer between the Fe foil and the sample is about 4~$\mu$m at $P=0$~GPa. The medium is NH$_3$BH$_3$.}
\label{fig:schematic}
\end{figure}

The measurement geometry is shown in Fig.~\ref{fig:schematic}. A diamond anvil cell was prepared with a composite gasket: an outer Re foil (350~$\mu$m thick, 400~$\mu$m hole) and an inner X-ray-transparent ceramic with a 60~$\mu$m culet and a 25~$\mu$m sample hole (expanding to $\sim$40~$\mu$m at high pressure). A Y foil and ammonia borane (NH$_3$BH$_3$), the latter serving as the hydrogen source, were loaded. A $^{57}$Fe sensor foil with an initial thickness of $\approx 0.5$~$\mu$m, a high-pressure thickness of $\approx 0.25$~$\mu$m (from NFS fits), and lateral dimensions at high pressure of $\approx 5\times 9$~$\mu$m$^2$ (confirmed by scanning the sensor profile against the NFS signal) was placed between the diamond anvil culet and the sample. The external field was applied perpendicular to the sample plane, parallel to the X-ray beam.

\begin{figure}[t]
  \includegraphics[width=\columnwidth ]{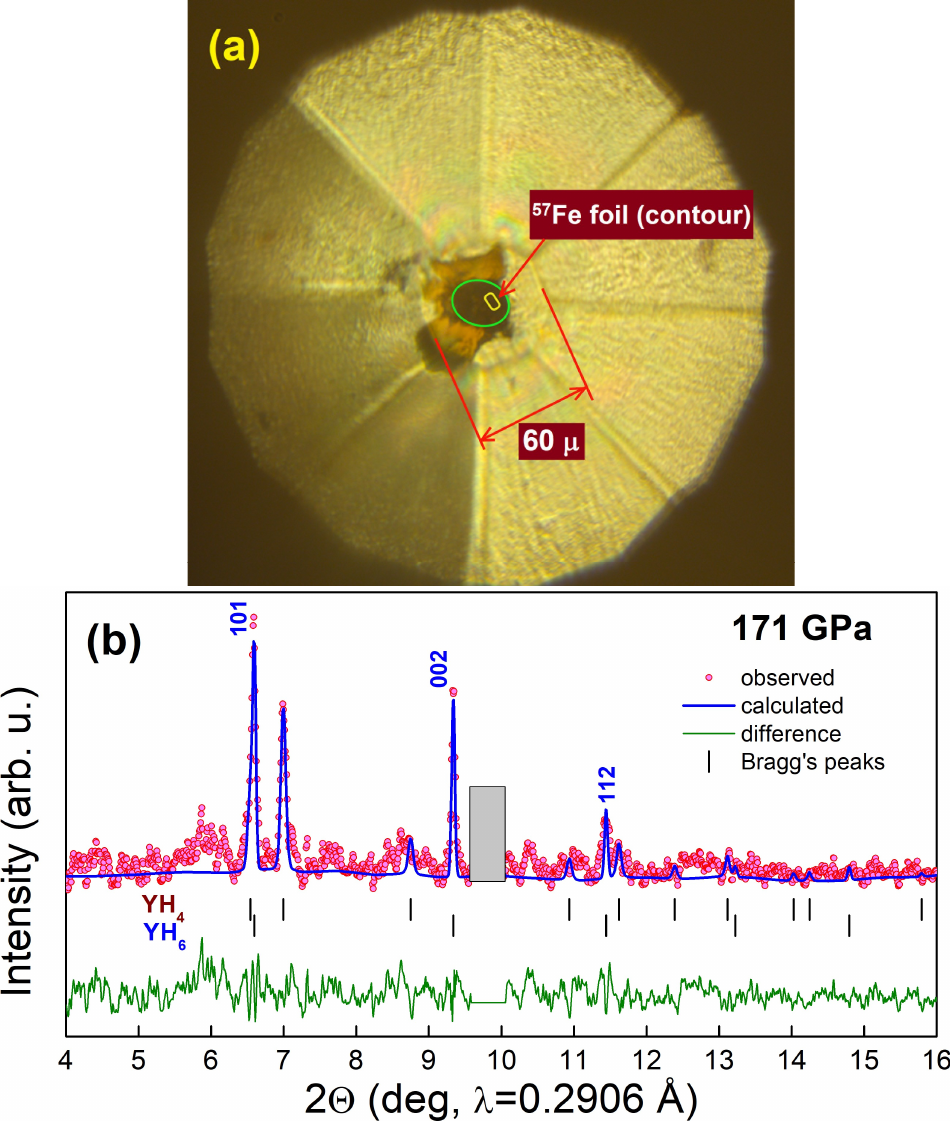}
\caption{(a) Sample microphotograph showing the Fe sensor location with respect to the laser-heated area. The approximate YH$_6$ location from x-ray mapping is indicated by green oval.  (b) XRD patterns of the YH$_x$ sample: phases YH$_6$, and YH$_4$, identified. No traces of Y metal were detected in or around YH$_6$ synthesis area.  After laser heating at 175~GPa the pressure decreased to 171~GPa.}
\label{fig:sample}
\end{figure}

After loading, pressure was increased to $\approx 175$~GPa and the sample was laser-heated from the side opposite the Fe sensor to avoid chemical reaction. The final pressure was 171~GPa. The sample microphotograph and the X-ray diffraction (XRD) pattern of the YH$_x$ cell are shown in Fig.~\ref{fig:sample}(a) and~\ref{fig:sample}(b), respectively. The XRD mapping reveals the superconducting phases YH$_4$ (I4/mmm) and YH$_6$ (Im-3m)~\cite{Troyan2021}; the phase composition and the sample dimensions inferred from these data are the inputs to the critical-state fits described below. The $^{57}$Fe sensor remained chemically inert and diamagnetic, as confirmed by room-temperature NFS spectra (isomer shift $0.69$~mm/s, consistent with the pressure dependence of the $\epsilon$-Fe isomer shift measured up to 241~GPa~\cite{Gavriliuk2023}).

Zero field was established at room temperature using a demagnetization cycle of the magnet: 1~T~$\rightarrow$~$-$0.1~T~$\rightarrow$~$+$0.05~T~$\rightarrow$~$-$0.02~T~$\rightarrow$~$+$0.005~T~$\rightarrow$~0~T. After cooling to 4.1~K in zero field (ZFC), NFS spectra were recorded on the virgin (ascending-field) branch. 

NFS experiments were performed at the P01 beamline of PETRA-III (DESY, Hamburg). The X-ray beam was monochromatized to 14.4125~keV with a 1.5~meV bandwidth and focused to $\approx 9\times 5.5$~$\mu$m$^2$ (FWHM). The DAC was mounted in a liquid-helium cryostat with a superconducting magnet (up to 6~T). The time-domain spectra were analyzed with the MOTIF package~\cite{Shvydko1999}, and the hyperfine field $B_{\rm Fe}$ was extracted with custom Python software based on the NEXUS engine~\cite{NEXUS}. The magnetic field value was accurate to $<1$~mT; the fit accuracy was $\sigma_B\approx 0.02$--$0.03$~T.

\section{Results}

\subsection{Measured $B_{\rm sensor}(B_a)$}

\begin{figure}[!t]
  \includegraphics[width=0.5\columnwidth ]{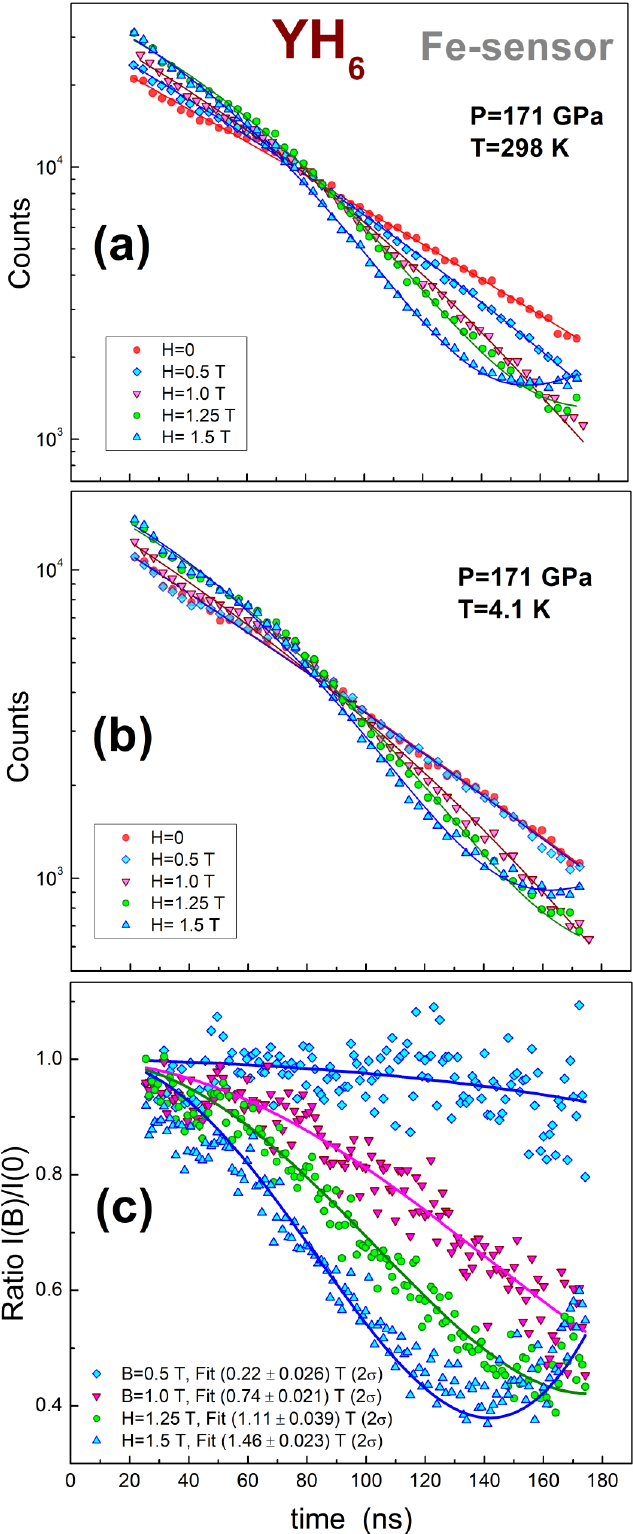}
\caption{Evolution of the NFS spectra of the $^{57}$Fe sensor: (a) at 298~K (normal state); (b) at 4.1~K (superconducting state), $P\approx 171$~GPa. Symbols: experiment; solid lines in (a), (b): fits using the MOTIF package. (c) Screened sensor field $B_{\rm sensor}$ versus applied field $B_a$, extracted from $I(B)/I(0)$ fits. The zero-field spectrum $I(0)$ was first fitted with an effective sensor thickness of approximately 0.25~$\mu$m. The thickness was then held fixed while each $I(B)/I(0)$ ratio was fitted by varying $B$ and a multiplicative intensity-scaling factor for $I(B)$.}
\label{fig:nfs}
\end{figure}

\begin{figure}[t]
  \includegraphics[width=\columnwidth]{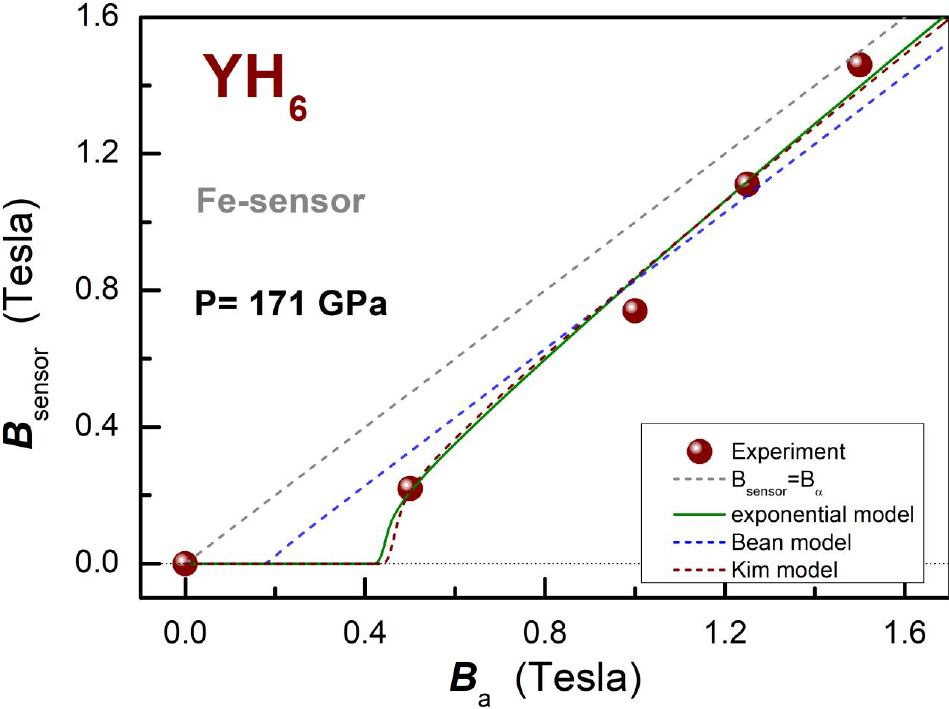}
\caption{Measured $B_{\rm sensor}$ vs.\ applied field $B_a$ at $T=4.1$~K and $P=171$~GPa (symbols), together with generalized critical-state model fits calculated with the SCS numerical method for a thin disk ($D=30$~$\mu$m, $d=2.5$~$\mu$m, $r/R=0.45$). The exponential and Kim models both reproduce the data, whereas the constant-$J_c$ Bean model does not. Parameters: Bean model $J_{c0}=(0.99\pm0.06)\times10^{11}$~A/m$^2$; Kim model $J_{c0}=(4.38\pm1.48)\times10^{11}$~A/m$^2$, $B_0=0.263\pm0.132$~T; exponential model $J_{c0}=(2.45\pm0.22)\times10^{11}$~A/m$^2$, $B_0=1.04\pm0.14$~T. The dashed grey line is $B_{\rm sensor}=B_a$.}
\label{fig:fits}
\end{figure}

Figure~\ref{fig:nfs} shows the evolution of the NFS spectra, at room temperature (normal state) and at 4.1~K (superconducting state), together with the screened sensor field $B_{\rm sensor}$ extracted from the $I(B)/I(0)$ ratio fits. The values of $B_{\rm sensor}$ as a function of the applied field $B_a$ at 4.1~K and 171~GPa are plotted in Fig.~\ref{fig:fits} (symbols). Three distinct regimes are predicted by SCS model: (i) a nearly perfectly shielding critical-state regime at low fields, $B_a\lesssim 0.4$~T; (ii) a penetration regime, $0.4\lesssim B_a\lesssim 1$~T, in which $B_{\rm sensor}$ rises steeply but remains below $B_a$; and (iii) a high-field regime in which $B_{\rm sensor}$ approaches and merges with the normal-state line, within experimental accuracy, at 1.5~T.

\subsection{SCS critical-state analysis}

Instead of fitting phenomenological formulas directly to the sensor field, we calculated the local induction $B_z(r)$ inside the sample numerically with the SCS method~\cite{Clem1994,Shantsev1999}, which solves the critical state of a thin superconducting disk. For the fits we used a sample diameter of 30~$\mu$m and a thickness of 2.5~$\mu$m, with the sensor position estimated at $r/R=0.45$ from the sample microphotograph (Fig.~\ref{fig:sample}). The finite beam size ($\approx 9\times 5.5$~$\mu$m$^2$) averages over approximately $r/R=0.45\pm0.15$. For each critical-state model---constant-$J_c$ Bean, Kim $J_c(H)=J_{c0}/(1+H/B_0)$, and exponential $J_c(H)=J_{c0}\exp(-H/B_0)$---the virgin branch $B_z(B_a)$ at $r/R=0.45$ was computed and fitted to the experimental fields $B_z\approx B_{\rm sensor}$. The details of the SCS implementation, including benchmark testing, are given in the End Matter.

The fitted parameters are: Bean model $J_{c0}=(0.99\pm0.06)\times10^{11}$~A/m$^2$; Kim model $J_{c0}=(4.38\pm1.48)\times10^{11}$~A/m$^2$, $B_0=0.263\pm0.132$~T; exponential model $J_{c0}=(2.45\pm0.22)\times10^{11}$~A/m$^2$, $B_0=1.04\pm0.14$~T. The Bean model retains a constant offset at high fields and fails to reproduce the rapid collapse of screening. Both the Kim and exponential models provide comparably good descriptions of the virgin curve. Their inferred $J_{c0}$ values differ, but the Kim uncertainty overlaps the exponential estimate; with the present five-point virgin curve we therefore do not reject either model. We instead treat them as two candidate generalized critical-state descriptions and use their spread as an estimate of model-dependent systematic uncertainty.

\section{Discussion}

A quantitative description of the virgin magnetization requires a critical-state model solved for the actual sample geometry, which is precisely what the SCS numerical method provides for a thin disk. The comparison in Fig.~\ref{fig:fits} shows that a field-dependent $J_c(B)$ is required: the constant-current Bean model cannot describe the high-field collapse of screening, whereas both the Kim and exponential laws can. These are generalized critical-state models in the sense that they introduce a field scale on which the local pinning current decreases; the SCS calculation then determines how this microscopic law is expressed in the measured sensor field.

The two retained models emphasize different aspects of the same physics. The exponential fit gives $J_{c0}=2.45\times10^{11}$~A/m$^2$ and $B_0=1.04$~T, describing a rapid weakening of pinning over a field scale of order 1~T. The Kim fit gives a larger zero-field value, $J_{c0}=4.38\times10^{11}$~A/m$^2$, but a much smaller characteristic field, $B_0=0.263$~T; over the measured field range the two parameter sets are strongly correlated and produce similar sensor fields. Exponential suppression of $J_c$ has been observed in strongly pinned superconductors such as cold-worked Nb--Zr wires~\cite{Fietz1964} and some high-$T_c$ cuprates~\cite{Xu1990}. In the present granular YH$_6$ sample it can be interpreted as progressive decoupling of Josephson-coupled grains. The Kim law remains a valid empirical candidate, however, and should not be discarded on the basis of the present virgin curve alone.

Across the two viable field-dependent models, the zero-field critical-current density lies in the range $J_{c0}\approx(2.45$--$4.38)\times10^{11}$~A/m$^2$, or $(2.45$--$4.38)\times10^7$~A/cm$^2$. These values are physically reasonable for a strongly pinned, granular critical state: they are at the upper end of typical 4.2~K critical-current densities but remain about two orders of magnitude below the YH$_6$ depairing limit~\cite{Eremets2022}. The critical densities reported by Eremets \textit{et al.}~\cite{Eremets2022} range from $\sim7\times10^{10}$~A/m$^2$ for H$_3$S at 100~K to $2.6\times10^{11}$~A/m$^2$ for YH$_6$, consistent with the lower part of our model range. Because the sample is granular, transport measurements return the intergranular critical-current density and therefore provide a lower bound on the local $J_{c0}$. The original transport estimate used a conservative cross-section of approximately $10\times50$~$\mu$m$^2$~\cite{Troyan2021}; using a compressed thickness of 2--3~$\mu$m and a laser-heated region of 2--4~$\mu$m raises the estimate to $\sim10^{11}$~A/m$^2$~\cite{Eremets2022}. Transport and NFS therefore agree in order of magnitude once the compressed geometry and granular current path are taken into account.

The granular structure also clarifies the physical interpretation of the fitted $J_c(B)$. The sample contains Josephson-coupled YH$_4$ and YH$_6$ grains, and the rapid field dependence of the fitted critical current can be interpreted as progressive decoupling of these weak links. For a granular superconductor, the relevant screening length is the Josephson penetration depth,
\[
\lambda_J=\sqrt{\frac{\hbar}{2e\mu_0dJ_c}},
\]
rather than the London penetration depth. Taking $J_c\approx2.45\times10^{11}$~A/m$^2$ and $d\approx2.5$~$\mu$m gives $\lambda_J\approx0.1$~$\mu$m, smaller than the typical grain size estimated from XRD peak broadening. This supports the use of the continuum SCS model on scales larger than $\lambda_J$. The NFS signal averages over the beam area and is weighted by the local field distribution; it primarily probes intra-granular shielding currents, whereas transport probes the inter-granular weak-link network.

Finally, we comment on the broader landscape of high-pressure magnetometry. The NFS method and optical quantum sensing techniques such as nitrogen-vacancy (NV) centers in diamond operate in complementary magnetic-field regimes~\cite{NVReview2026}.  NV magnetometry provides exceptional sensitivity and sub-micron spatial resolution at low fields ($\lesssim 0.1$~T)~\cite{Doherty2014} and has been extended to megabar pressures for imaging the Meissner effect in hydride superconductors~\cite{Bhattacharyya2024,Chen2026,Hilberer2023}. However, NV optically detected magnetic resonance becomes increasingly challenging above $\sim 0.12$~T because the large Zeeman splitting shifts the resonances out of the microwave range. NFS, by contrast, provides robust quantitative field measurements in the range 0.1--10~T, exactly the window in which the critical-state response of YH$_6$ unfolds. The two techniques are therefore highly complementary; an overlap below 0.1~T would allow cross-calibration and combined studies of vortex dynamics across the full field range. 

Taken together, our results resolve the controversy surrounding the interpretation of M\"ossbauer sensor experiments on superhydrides~\cite{Troyan2016,Hirsch2021,Prozorov2022}. The observed flux expulsion is not the equilibrium Meissner effect but a pinned critical state. The virgin data establish the need for a field-dependent $J_c(B)$ and strongly disfavor the constant-current Bean limit, but they do not uniquely select between the Kim and exponential laws. Hysteresis loops, spatially resolved field maps, or improved constraints on the sensor position would provide stronger model discrimination.

\section{Conclusions}

We have demonstrated that a $^{57}$Fe M\"ossbauer sensor inside a diamond anvil cell can measure the local magnetic field in a polyhydride superconductor with a resolution of 0.02--0.03~T. For YH$_6$ at 171~GPa and 4.1~K, the $B_{\rm sensor}(B_a)$ data are described by generalized field-dependent critical-state models evaluated with the SCS numerical method for a thin disk. The exponential fit gives $J_{c0}=(2.45\pm0.22)\times10^{11}$~A/m$^2$ and $B_0=1.04\pm0.14$~T; the comparably good Kim fit gives $J_{c0}=(4.38\pm1.48)\times10^{11}$~A/m$^2$ and $B_0=0.263\pm0.132$~T. The constant-$J_c$ Bean model fails on the virgin curve. Thus the observed flux expulsion is due to vortex pinning in a field-dependent critical state, not the reversible Meissner effect, while the exact microscopic pinning law remains model-dependent. The methodology can be extended to other high-pressure superconductors and to hysteresis and magnetization measurements.

\begin{acknowledgments}
This work was supported by the National Key Research and Development Program of China (grant 2023YFA1608900, subproject 2023YFA1608903).  V.V.Struzhkin and I.A.Troyan thank the NSFC International Collaboration Fund for Creative Research Teams, grant No. W2641002, for financial support.  Support from the Russian Science Foundation (RSF grant 27-12-00323) is also acknowledged. The work performed by A. G. Gavriliuk was carried out within the framework of the State Assignment of the Institute for Nuclear Research, Russian Academy of Sciences (INR RAS), and the experimental samples were prepared at its shared research facility.   We thank the staff of the P01  and P02.2  beamlines at PETRA-III (DESY, Hamburg) for technical support. This work was also supported by the European Research Council (ERC Advanced Grant 267777).
\end{acknowledgments}

\clearpage

\clearpage
\section*{End Matter}

\subsection*{S1. Shantsev--Clem--Sanchez numerical implementation}

The solver follows the critical-state construction of Shantsev \textit{et al.}~\cite{Shantsev1999}, using the original field-current relations and, in particular, Eqs.~(6)--(11) for the virgin branch. The physical equations were not changed. The differences are computational and were introduced to make the published algorithm stable and fast enough for fitting the YH$_x$ data.

First, the evaluation of Eq.~(9) was vectorized over the radial grid instead of using nested scalar loops. Second, the endpoint singularity at the flux front was removed by the substitution
\[
u=\sqrt{r'^2-a^2},
\]
after which the transformed integral was evaluated on a uniform $u$-grid. Third, the flux-front bisection was performed with a fast algebraic preview, followed by one final converged evaluation of the original integral. Finally, adaptive damping and warm starts from the previous field point were used for field-dependent $J_c(B)$ models; the Bean branch uses the analytic Clem--S\'anchez solution directly. These changes affect only convergence speed and discretization error, not the critical-state equations.

For the benchmark case of constant \(J_c\), the numerical solution was validated against the exact Clem--S\'anchez analytical solutions for a thin disk~\cite{Clem1994,Shantsev1999}. We varied the radial grid from 100 to 800 points and tightened the convergence tolerance until the flux-front and field profiles changed by less than \(10^{-6}\) of the applied field. All Bean-model benchmarks agreed with the analytical results to better than 1\%. For the Kim and exponential models, where no corresponding analytical solution exists, convergence was tested by increasing the radial resolution and tightening the tolerance until the fitted parameters were stable.

The fitting implementation was optimized with coding assistance from Kimi 3.0 and DeepSeek. The solver, CLI fitting software, and the data  are available from Zenodo~\cite{SCSSoftware}.

\subsection*{S2. Sensitivity to sample geometry and sensor position}

The fits in the main text use the  geometry $D=30$~$\mu$m, $d=2.5$~$\mu$m, and $r/R=0.45$. To assess the robustness of the extracted exponential-model parameters, we varied these quantities over the estimated uncertainty ranges $D=30\pm5$~$\mu$m, $d=2.5\pm0.5$~$\mu$m, and $r/R=0.45\pm0.15$. The diameter variation has a negligible effect in the thin-disk limit. The thickness changes $J_{c0}$ approximately inversely, whereas the radial sensor position remains the dominant geometric uncertainty. The propagated parameter uncertainties in the main text were obtained from $\sigma_B=0.020$~T using finite-difference sensitivity analysis. 
The sample is not exactly circular, which introduces a further uncertainty that is difficult to address in actual experiments. However, we have followed a simplified protocol similar to that of recent studies~\cite{Minkov2022,Minkov2023,Troyan2016,Hirsch2021,Prozorov2022} and hope that our results will be refined in future experiments with better geometrical constraints and tighter X-ray focusing. The observed insensitivity of the fit results to the sample diameter provides indirect support for the idealized circular geometry assumed in the analysis.

\begin{table}[h!]
\centering
\caption{Sensitivity of exponential-model parameters to geometric uncertainties. Nominal values: $D=30$~$\mu$m, $d=2.5$~$\mu$m, $r/R=0.45$.}
\label{tab:sensitivity}
\scriptsize
\setlength{\tabcolsep}{2pt}
\begin{tabular}{lccccc}
\hline\hline
Parameter & Value & $J_{c0}$ ($10^{11}$~A/m$^2$) & $\Delta J_{c0}$ (\%) & $B_0$ (T) & $\Delta B_0$ (\%) \\
\hline
Nominal                 & ---        & 2.45 & ---    & 1.04 & ---    \\
$D$                     & 25~$\mu$m  & 2.45 & $-0.0$ & 1.04 & +0.2   \\
$D$                     & 35~$\mu$m  & 2.45 & $-0.0$ & 1.04 & +0.2   \\
$d$                     & 2.0~$\mu$m & 3.06 & +25.0  & 1.04 & +0.0   \\
$d$                     & 3.0~$\mu$m & 2.05 & $-16.2$& 1.03 & $-0.5$ \\
$r/R$                   & 0.30       & 1.73 & $-29.3$& 1.01 & $-2.4$ \\
$r/R$                   & 0.60       & 3.38 & +37.9  & 1.21 & +16.8  \\
\hline\hline
\end{tabular}
\end{table}

\subsection*{S3. Sensor standoff}

The Fe sensor is displaced by an estimated $h=1$--2~$\mu$m from the sample plane. A simple standoff estimate can be obtained by approximating the screened disk as an effective current distribution whose self-field decays approximately as $1/(1+h/R)^2$ once the sensor is displaced from the plane. Defining the shielding deficit
\[
D(B_a)=B_a-B_{\rm sensor}(0),
\]
the field measured at height $h$ is approximately
\[
B_{\rm sensor}(h)\simeq B_a-\frac{D(B_a)}{(1+h/R)^2}.
\]
Thus, the systematic upward shift relative to the in-plane interpretation is
\[
\Delta B_{\rm sensor}(h)
\simeq D(B_a)\left[1-\frac{1}{(1+h/R)^2}\right].
\]

For $R=15$~$\mu$m and the measured in-plane fields, the estimates are given in Table~\ref{tab:standoff-simple}. The zero-field point has no standoff correction because there is no screening self-field.

\begin{table}[h!]
\centering
\caption{Simple sensor-standoff estimates using $B_{\rm sensor}=[0,0.22,0.77,1.11,1.46]$~T at $B_a=[0,0.50,1.00,1.25,1.50]$~T. The equivalent geometric uncertainty assumes $h=1.5\pm0.5$~$\mu$m.}
\label{tab:standoff-simple}
\scriptsize
\setlength{\tabcolsep}{2pt}
\resizebox{\columnwidth}{!}{%
\begin{tabular}{cccccc}
\hline\hline
$B_a$ (T) & $B_{\rm sensor}(0)$ (T) & $B_a-B_{\rm sensor}(0)$ (T) & $\Delta B(1~\mu{\rm m})$ (T) & $\Delta B(2~\mu{\rm m})$ (T) & $\sigma_{B,\mathrm{geom}}$ (T) \\
\hline
0    & 0.00 & 0.00 & 0.000 & 0.000 & 0.000 \\
0.50 & 0.22 & 0.28 & +0.034 & +0.062 & 0.014 \\
1.00 & 0.77 & 0.23 & +0.028 & +0.051 & 0.012 \\
1.25 & 1.11 & 0.14 & +0.017 & +0.031 & 0.007 \\
1.50 & 1.46 & 0.04 & +0.005 & +0.009 & 0.002 \\
\hline\hline
\end{tabular}%
}
\end{table}

The standoff therefore produces a positive model systematic, largest at intermediate fields. If the sensor displacement were known accurately, an exact off-plane Biot--Savart calculation could be included in the forward model. Since only the approximate range $h=1$--2~$\mu$m is known here, the simple estimate above is sufficient and the quoted standoff contribution should be treated as a systematic uncertainty rather than an additional random field uncertainty.

\subsection*{S4. Quality factor of M\"ossbauer isotopes for magnetic-field sensing}

The ability of a M\"ossbauer isotope to measure a small magnetic field $B$ is determined by the Zeeman splitting of the nuclear level relative to the natural linewidth $\Gamma$,
\begin{equation}
\frac{\Delta E(B)}{\Gamma}=\frac{g_n\mu_NB}{\Gamma}=QB,
\end{equation}
where $g_n=\mu_e/I_e$ is the nuclear $g$-factor of the M\"ossbauer (excited) level. The ratio $Q=g_n\mu_N/\Gamma$ is a quality factor: it is the splitting per unit field expressed in natural linewidths, equivalently the ratio of the nuclear magnetic moment (per unit spin) to the linewidth. A larger $Q$ gives a larger splitting for a given field and hence a higher signal-to-noise ratio for a given counting time.

The common isotopes cluster at $Q\approx 0.5$--$0.7$~T$^{-1}$: for $^{57}$Fe, $^{119}$Sn, $^{151}$Eu, and $^{161}$Dy the splitting reaches one natural linewidth at $B_\Gamma\approx 1.4$--$2.2$~T. Their intrinsic Zeeman sensitivity is therefore comparable, and none offers a dramatic gain in $Q$ over $^{57}$Fe. The practical resolution achieved in NFS ($\sigma_B\approx 0.02$~T for $^{57}$Fe with 15--30~min counting) is set by counting statistics, inhomogeneous broadening, and the length of the measurement time window rather than by $\Gamma$, so that a moderate difference in $Q$ translates into only a modest gain in sensitivity. $^{237}$Np stands apart with $Q=3.1$~T$^{-1}$ ($B_\Gamma=0.32$~T), making it the most sensitive choice for the sub-tesla range; its use is restricted, however, by actinide handling, sample availability, and the need for specialized beamline optics. The larger magnetic moments of $^{151}$Eu and $^{119}$Sn are partially offset by their short level lifetimes ($\tau\approx 14$~ns and 26~ns, respectively, compared with 141~ns for $^{57}$Fe), which broaden the lines and require fast detectors.

The choice of isotope is thus governed less by the quality factor than by experimental considerations: the photon flux available at the transition energy, the resonance cross section and recoilless fraction, the natural abundance, and chemical compatibility with the sample and the DAC. On these grounds $^{57}$Fe and $^{119}$Sn remain the most practical choices for the field range of the present experiment (0.1--10~T), whereas $^{237}$Np would be preferable for measurements below $\sim 0.1$~T.


\begin{thebibliography}{99}
\bibitem{Drozdov2015} A.~P.~Drozdov, M.~I.~Eremets, I.~A.~Troyan, V.~Ksenofontov, and S.~I.~Shylin, Nature \textbf{525}, 73 (2015).
\bibitem{Somayazulu2019} M.~Somayazulu, M.~Ahart, A.~K.~Mishra, Z.~M.~Geballe, M.~Baldini, Y.~Meng, V.~V.~Struzhkin, and R.~J.~Hemley, Phys. Rev. Lett. \textbf{122}, 027001 (2019).
\bibitem{Drozdov2019} A.~P.~Drozdov, P.~P.~Kong, V.~S.~Minkov, S.~P.~Besedin, M.~A.~Kuzovnikov, S.~Mozaffari, L.~Balicas, F.~F.~Balakirev, D.~E.~Graf, V.~B.~Prakapenka, E.~Greenberg, D.~A.~Knyazev, M.~Tkacz, and M.~I.~Eremets, Nature \textbf{569}, 528 (2019).
\bibitem{Minkov2022} V.~S.~Minkov, S.~L.~Bud'ko, F.~F.~Balakirev, V.~B.~Prakapenka, S.~Chariton, R.~J.~Husband, H.-P.~Liermann, and M.~I.~Eremets, Nat. Commun. \textbf{13}, 3194 (2022).
\bibitem{Minkov2023} V.~S.~Minkov, V.~Ksenofontov, S.~L.~Bud'ko, E.~F.~Talantsev, and M.~I.~Eremets, Nat. Phys. \textbf{19}, 1293 (2023).
\bibitem{Bhattacharyya2024} P.~Bhattacharyya, W.~Chen, X.~Huang, S.~Chatterjee, B.~Huang, B.~Kobrin, Y.~Lyu, T.~J.~Smart, M.~Block, E.~Wang, Z.~Wang, W.~Wu, S.~Hsieh, H.~Ma, S.~Mandyam, B.~Chen, E.~Davis, Z.~M.~Geballe, C.~Zu, V.~Struzhkin, R.~Jeanloz, J.~E.~Moore, T.~Cui, G.~Galli, B.~I.~Halperin, C.~R.~Laumann, and N.~Y.~Yao, Nature \textbf{627}, 73 (2024).
\bibitem{Chen2026} Y.~Chen, J.~Wen, Z.-X.~He, J.-W.~Fan, X.-Y.~Pan, C.~Ji, H.~Gou, X.~Yu, L.~Chen, and G.-Q.~Liu, Proc. Natl. Acad. Sci. USA \textbf{123}, e2536915123 (2026).
\bibitem{Troyan2016} I.~Troyan, A.~Gavriliuk, R.~Rüffer, A.~Chumakov, A.~Mironovich, I.~Lyubutin, D.~Perekalin, A.~Drozdov, and M.~Eremets, Science \textbf{351}, 1303 (2016).
\bibitem{Gerdau1985} E.~Gerdau, R.~R\"uffer, H.~Winkler, W.~Tolksdorf, C.~P.~Klags, and J.~P.~Hannon, Phys. Rev. Lett. \textbf{54}, 835 (1985).
\bibitem{Hastings1991} J.~B.~Hastings, D.~P.~Siddons, U.~van B\"urck, R.~Hollatz, and U.~Bergmann, Phys. Rev. Lett. \textbf{66}, 770 (1991).
\bibitem{Wortmann2002} G.~Wortmann, K.~Rupprecht, and H.~Giefers, Hyperfine Interact. \textbf{144/145}, 103 (2002).
\bibitem{Lubbers1999} R.~L\"ubbers, G.~Wortmann, and H.~F.~Gr\"unsteudel,  R.~R\"uffer, O.~Leupold, and J.~Zukowski, Hyperfine Interact. \textbf{120}, 49 (1999).
\bibitem{Sturhahn2004} W.~Sturhahn, J. Phys.: Condens. Matter \textbf{16}, S497 (2004).
\bibitem{Roehlsberger2004} R.~R\"ohlsberger, \textit{Nuclear Condensed Matter Physics with Synchrotron Radiation} (Springer, Berlin, 2004).
\bibitem{Gavriliuk2005} A.~G.~Gavriliuk, V.~V.~Struzhkin, I.~S.~Lyubutin, M.~Y.~Hu, and H.-K.~Mao,   JETP Lett. \textbf{82}, 224 (2005) .
\bibitem{Gavriliuk2018} A.~G.~Gavriliuk, V.~V.~Struzhkin, A.~A.~Mironovich, I.~S.~Lyubutin, I.~A.~Troyan, P.~Chow, and Y.~Xiao,   JETP Lett. \textbf{107}, 247 (2018).
\bibitem{Gavriliuk2023} A.~G.~Gavriliuk, V.~V.~Struzhkin, S.~N.~Aksenov, A.~A.~Mironovich, I.~A.~Troyan, A.~G.~Ivanova, and I.~S.~Lyubutin,  JETP Lett. \textbf{117}, 126 (2023).
\bibitem{Li2006} J.~Li, W.~Sturhahn, J.~M.~Jackson, V.~V.~Struzhkin, J.-F.~Lin, J.~Zhao, H.-K.~Mao, and G.~Shen, Phys. Chem. Miner. \textbf{33}, 575 (2006).
\bibitem{Bi2015} W.~Bi, J.~Zhao, J.-F.~Lin, Q.~Jia, M.~Y.~Hu, C.~Jin, R.~Ferry, W.~Yang, V.~Struzhkin, and E.~E.~Alp,  J. Synchrotron Radiat. \textbf{22}, 760 (2015).\bibitem{Troyan2021} I.~A.~Troyan, D.~V.~Semenok, A.~G.~Kvashnin, A.~V.~Sadakov, O.~A.~Sobolevskiy, V.~M.~Pudalov, A.~G.~Ivanova, V.~B.~Prakapenka, E.~Greenberg, A.~G.~Gavriliuk, I.~S.~Lyubutin, V.~V.~Struzhkin, A.~Bergara, I.~Errea, R.~Bianco, M.~Calandra, F.~Mauri, L.~Monacelli, R.~Akashi, and A.~R.~Oganov, Adv. Mater. \textbf{33}, 2006832 (2021).
\bibitem{Hirsch2021} J.~E.~Hirsch and F.~Marsiglio, Physica C \textbf{587}, 1353896 (2021).
\bibitem{Bean1962} C.~P.~Bean, Phys. Rev. Lett. \textbf{8}, 250 (1962).
\bibitem{Prozorov2022} R.~Prozorov and S.~L.~Bud'ko,  J.  Supercond. Nov. Magn. \textbf{35}, 2615 (2022).
\bibitem{Clem1994} J.~R.~Clem and A.~S\'anchez, Phys. Rev. B \textbf{50}, 9355 (1994).
\bibitem{Shantsev1999} D.~V.~Shantsev, A.~A.~Pukha, and Yu.~A.~Genenko, Phys. Rev. B \textbf{60}, 13112 (1999).
\bibitem{Mikheenko1993} P.~N.~Mikheenko and Yu.~E.~Kuzovlev, Physica C \textbf{204}, 229 (1993).
\bibitem{Zhu1993} J.~Zhu, J.~Mester, J.~Lockhart, and J.~Turneaure, Physica C \textbf{212}, 216 (1993).
\bibitem{Kim1962} Y.~B.~Kim, C.~F.~Hempstead, and A.~R.~Strnad, Phys. Rev. Lett. \textbf{9}, 306 (1962).
\bibitem{Fietz1964} W.~A.~Fietz, M.~R.~Beasley, J.~Silcox, and W.~W.~Webb, Phys. Rev. \textbf{136}, A335 (1964).
\bibitem{Xu1990} M.~Xu, D.~Shi, and R.~F.~Fox, Phys. Rev. B \textbf{42}, 10773(R) (1990).
\bibitem{Shvydko1999} Y.~V.~Shvyd'ko, Phys. Rev. B \textbf{59}, 9132 (1999).
\bibitem{NEXUS} NEXUS---Nuclear Elastic X-ray scattering Universal Software, DESY, \url{https://fs-mcp.pages.desy.de/nuclear-nexus/}.
\bibitem{Eremets2022} M.~I.~Eremets, V.~S.~Minkov, A.~P.~Drozdov, P.~P.~Kong, S.~I.~Shylin, S.~L.~Bud'ko, R.~Prozorov, F.~F.~Balakirev, D.~Sun, S.~Mozaffari, and L.~Balicas,  J.~Supercond.~Nov.~Magn. \textbf{35}, 965 (2022).
\bibitem{NVReview2026} Kin~On~Ho and Sen~Yang, npj Adv. Manuf. \textbf{3}, 33 (2026).
\bibitem{Doherty2014} M.~W.~Doherty, V.~V.~Struzhkin, D.~A.~Simpson, L.~P.~McGuiness,  Y.~Meng,  A.~Stacey, T.~J.~Karle, R.~J.~Hemley,  N.~B.~Manson,  L.~C.~L.~Hollenberg,  S.~Prawer, Phys. Rev. Lett.  \textbf{112}, 047601(2014).
\bibitem{Hilberer2023} A.~Hilberer, L.~Toraille, C.~Dailledouze, M.-P.~Adam, L.~Hanlon, G.~Weck, M.~Schmidt, P.~Loubeyre, and J.-F.~Roch, Phys. Rev. B \textbf{107}, L220102 (2023).
\bibitem{SCSSoftware} SCS solver,  fitting software, and the data available from  Zenodo, [DOI:10.5281/zenodo.24356881; DOI:10.5281/zenodo.24356833].
\end{thebibliography}
\end{document}